\documentclass[12pt]{iopart}

\usepackage{iopams}

\usepackage{mathrsfs}
\usepackage{bbm}
\usepackage{stmaryrd}
\usepackage{bm}
\usepackage{harvard}
\usepackage{pfnote}

\usepackage{amsfonts}
\usepackage{graphicx}        

\begin{document}
\title[Contributions of plasma physics to chaos and nonlinear dynamics]{Contributions of plasma physics to chaos and nonlinear dynamics}

\author{D~F Escande}
\address{Aix-Marseille Universit\'{e}, CNRS, PIIM, UMR 7345,
   case 321, campus Saint-J\'er\^ome, FR-13013 Marseille, France}
\eads{\mailto{dominique.escande@univ-amu.fr}}
\begin{abstract}
This review focusses on the contributions of plasma physics to
chaos and nonlinear dynamics bringing new methods which are or can be
used in other scientific domains. It starts with the development of the
theory of Hamiltonian chaos, and then deals with order or quasi order,
for instance adiabatic and soliton theories. It ends with a shorter account
of dissipative and high dimensional Hamiltonian dynamics, and of quantum
chaos. Most of these contributions are a spin-off of the research on
thermonuclear fusion by magnetic confinement, which started in the fifties.
Their presentation is both exhaustive and compact.
[\today]
\end{abstract}



\par \medskip
%
\maketitle
\section{Introduction}

\label{intro}

In the second half of the twentieth century, the magnetic confinement
of charged particles in magnetic bottles was studied to heat plasmas to
the high temperatures required for controlled thermonuclear fusion. The
basic idea was to put particles in magnetic fields whose lines wind upon
magnetic surfaces. Rapidly, it turned out that these lines are described
by low dimensional Hamiltonian mechanics, forcing physicists to revisit
classical mechanics and to get acquainted with the related results
mathematicians started to develop in the first half of the twentieth
century\footnote{\cite{Morrison2000} makes at length the point that
plasma physics re-ignited research in classical dynamics.}. Unfortunately,
though very useful, these results were not sufficient to deal
with magnetic confinement. Fortunately, the development of computers and of numerical
calculations enabled them to visualize dynamics, to compute their features,
and to back up the development of non rigorous approaches. While textbooks
were full of examples with regular motion, chaos turned out to be an
ubiquitous behaviour of classical mechanics!

A parallel endeavour in the second half of the twentieth century
occurred in the context of accelerator physics, with the aim to drive
particles to the high energies necessary for studying the physics of
elementary particles. This means avoiding chaos and not describing it,
which limits the scope of studies on this topic. In contrast, the
quest for magnetic fusion requires also the study of chaos itself, in order to understand chaotic transport. Indeed, this issue is
present in very different parts of the research on magnetic fusion:
magnetic field line topology, dynamics of particles in magnetic fields,
turbulent transport, radiofrequency heating, ray dynamics, etc\footnote{Chaos is at work in particular in anomalous transport of fusion plasmas, when heating magnetized plasmas by cyclotronic waves, and in the deep penetration of lower hybrid rays in such plasmas.}... This
broad scope of problems naturally triggered a series of contributions
to nonlinear dynamics and chaos. These contributions are the topic of
the present topical review. In order to limit its length, \emph{it
focusses on the contributions bringing new methods which are or can
be used in other scientific domains}. These contributions are of several
types. There are those of plasma physicists and those inspired by plasma
physics to scientists of other fields of science, mathematicians and
astronomers in particular. There are also those of plasma physicists
initially working on chaos and nonlinear dynamics, who then decided to
dedicate their later research to this exciting and challenging topic.
Plasma physics may not vindicate their whole successive work, but at
least the initial impetus it provided them.

This topical review starts with the development of the theory of Hamiltonian chaos in the context of plasmas (section \ref{Hacha}). Indeed, this was a turning point in plasma physics, because this development uncovered the then mysterious dynamics underlying phenomena traditionally tackled by statistical approaches; in particular, thanks to images provided by numerical calculations. This induced a strong interest in the whole plasma community, even by non practitioners of chaos.

The dynamics of complex systems generally exhibits a mixture of ordered and chaotic motion. Furthermore, a chaotic orbit needs a sufficient time to really look chaotic. Therefore, there is a large part of the nonlinear dynamics of plasmas which deals with order or quasi order, for instance adiabatic theory. This is the topic of the second part of this topical review (section \ref{Qord}). It deals, in particular, with soliton theory, which is an early landmark of the contributions of plasma physics to nonlinear dynamics.

The following parts are shorter and deal with high dimensional Hamiltonian dynamics (section \ref{hdH}), dissipative dynamics (section \ref{disdyn}), quantum chaos (section \ref{Qcha}), and possible extensions (section \ref{Pex}).

Writing the history of the contributions of plasma physics to chaos and nonlinear dynamics is a delicate endeavour. Indeed, the beginning of this history corresponds to a literature whose access is often not easy, in particular for linguistic reasons, while the more modern part suffers from the present deluge of publications where interesting results may be drowned. In order to limit bias, a series of colleagues listed in the acknowledgements kindly provided me with their views on the topic of this topical review. This considerably enriched its initial scope, but it may still be incomplete.

Telling the same story with equations and figures would require a book. The present compact and hopefully close to exhaustive paper is tailored for the plasma physics community, and especially the fusion one, where most concepts underlying this story are well-known. In any event, more details can be obtained in the about 250 quoted references with a simple click. However, in order to help non specialists to access more easily to the basics of the main issues, references to sections of review papers and textbooks are provided in the corresponding sections of this review and are indicated with ``Pedagogy".

\section{Hamiltonian chaos}
\label{Hacha}
\subsection{How did the story start?}

After the second world war, classical mechanics was not fashionable at all among physicists. Why did plasma physics happen to contribute to this field? It is because many plasma physicists were involved in the development of thermonuclear fusion by magnetic confinement. The story can be told as follows.

In 1950, Spitzer invents the stellarator, and in 1951 Sakharov and Tamm invent the tokamak. While the principle of the latter comes with a regular magnetic field, this cannot be taken for granted for the former, a serious issue for a magnetic configuration meant to be a magnetic bottle! This forces theoreticians to start the still on-going study of the nature of magnetic field lines of stellarators. Even if a stellarator looks like a figure 8 as Spitzer's one, topologically it is a torus. Therefore, the regularity of its magnetic field lines can be checked by looking at their successive intersections with the surface of section defined by a given toroidal angle. A massless charged particle streaming freely along a given magnetic field line crosses such a surface periodically in time. By analogy, it is natural to consider this line as an orbit parameterized by a time which is the toroidal angle. Then, studying the nature of magnetic field lines boils down to the study of the nature of orbits in a torus, and their successive intersections build the so-called Poincar\'e map of their dynamics\footnote{This technique is also used experimentally. The first instance is the mapping of the magnetic field lines of the model-C Stellarator \cite{Sinclair1970}.}. Magnetic flux conservation makes such a map area-preserving, implying a relationship between magnetic field lines and Hamiltonian systems.

While in 1952 Kruskal\footnote{Kruskal is quoted several time in this topical review. Indeed, he made essential contributions to nonlinear dynamics and chaos. He was also quite influential. In particular in astrophysics, as can be seen in the acknowledgements of the H\'enon-Heiles paper where he is thanked \cite{Henon1964}. Indeed, this famous work was performed while Michel H\'enon was in Princeton. It describes the non-linear motion of a star around a galactic center where the motion is restricted to a plane, and uncovers this motion can be chaotic.}
introduces and iterates area preserving maps for stellarator magnetic field lines \cite{Kruskal1952}, they are fully recognized as Hamiltonian systems only ten years later by fusion physicists \cite{Kerst1962,Gelfand1962,Morozov1966}. However, the general and explicit Hamiltonian description for field lines is provided by Boozer even two decades after \cite{Boozer1981}\footnote{Boozer is a plasma physicist whose paper is in \textit{Physics of Fluids} which published both fluid and plasma
papers at that time.}. Two years later, a more fundamental description is given by two other plasma physicists, Cary and Littlejohn, in terms of an action principle that depends on the vector potential\footnote{Reference \cite{Pina1988} describes how to implement this technique in concrete cases and illustrates its large flexibility.} \cite{Cary1983}\footnote{This action principle is present implicitely in \cite{Morozov1966} where a zero gyro-radius limit of the particle action was taken.}. In 1986, Els\"asser completes the picture by showing the equivalence of changes of gauge and of canonical variables \cite{Elsasser1986}.

Having a Hamiltonian description of magnetic field lines is nice, but when coming to the numerical calculation of Poincar\'e maps, the integration of orbits from differential equations is a formidable task for the computers of the sixties! This motivates physicists to derive explicit area preserving maps corresponding to a full step of the Poincar\'e map. The paradigm of such maps is the
standard map, also called Chirikov-Taylor map. This map appeared first in 1960 in the context of electron dynamics in the microtron\footnote{According to reference 2 of \cite{Melekhin1975}, it appeared ten years earlier in Kolomenskii's PhD thesis at the Lebedev Institute.} \cite{Kolomenskii1960}, a type of particle accelerator concept originating from the cyclotron in which the accelerating field is not applied through large D-shaped electrodes, but through a linear accelerator structure. This map was independently proposed and numerically studied in a magnetic fusion context by Taylor
\footnote{Taylor writes : ``At the time I was interested in [the standard map], I had just obtained an early desk calculator which
could be programmed using three stored addresses $(x,y,z)$, but more importantly it
could be connected to a mechanical plotter! One lunch time I set it up to iterate the
equations above, and to plot the successive points - which it did at about two per second!
When I returned from lunch there was the first picture of regular and chaotic
regions. The others followed later.
I did not consider this work suitable for publication, but I did include the figures in
the Culham Progress report for that year (1968-9). This was noticed and taken up by
others, notably Froeschl\'e \cite{Froeschle1970} who published the figures (with acknowledgment) and by Stix, who included them in his lectures. So the model became quite well known." \cite{Taylor2015}.} in 1968, and by Chirikov\footnote{Chirikov is quoted repeatedly in this paper for contributions in many different problems of nonlinear dynamics and chaos. A summary of his main contributions can be found in \cite{Bellissard1999}, published in a special issue of Physica D in his honor.} in 1969 (also for particle accelerators) \cite{Chirikov1969}. The latter also recovers this map on linearizing the motion in the vicinity of a separatrix (section 4.4 of \cite{Chirikov1979}).

\subsection{Transition to chaos in Hamiltonian systems}
\label{TCHSys}

In the sixties, numerical simulations make visible that the phase-space of a typical 1.5 degree-of-freedom Hamiltonian system has intertwined zones of regular and chaotic orbits, a fact known by mathematicians for decades. For plasma physicists it is important to know how broad are chaotic domains\footnote{Then called ``stochastic domains". The change of name is motivated by several reasons. A simple one comes from the inspection of the time evolution of two nearby orbits at initial time. In a chaotic system they diverge exponentially, but not in a stochastic one.}, and the threshold where such domains connect to make sizeable chaotic ``seas" for systems with a control parameter. The latter issue leads Chirikov to derive in 1959 an approximate criterion \cite{Chirikov1959,Chirikov1979}, described below,  whose use becomes rapidly ubiquitous. This criterion works for Hamiltonians, which are the sum of an integrable part and of a perturbation written in terms of the action-angle variables of the integrable part. Kolmogorov-Arnold-Moser (KAM) theory \cite{Kolmogorov1954,Moser1962,Arnold1963a} reveals that the extension of chaotic domains is bounded by KAM tori\footnote{The quotations of \cite{Arnold1963b} and \cite{Moser1968} show that plasma physics is also a source of inspiration for these mathematicians.}. Therefore, a more rigorous approach to the estimate of the width of chaotic domains goes through the estimate of the threshold of break-up of KAM tori. Greene\footnote{Greene is quoted repeatedly in this paper for contributions in many different problems of nonlinear dynamics and chaos. A summary of his main contributions can be found in \cite{Morrison2008}.} achieves this in 1979 in a very accurate way through a criterion \cite{Greene1979}, described below, which is explicitly implemented for the standard map\footnote{Scalings in \cite{Greene1968} anticipate this result.}. As a result, the transition to chaos is then described in two complementary ways using Hamiltonians and maps. The next two subsections tell the corresponding story. The first one describes the Hamiltonian approach which started first historically.

\subsubsection{Working with Hamiltonians}

\paragraph{Resonance-overlap criterion}
\label{ResOver}

In 1959, Chirikov deals with the confinement of charged particles in
magnetic mirror traps. He focusses on resonances between the Larmor
rotation of charged particles and their slow oscillations along the
lines of force. In agreement with numerical simulations, he hypothesizes
that when neighboring resonances overlap, there is a complete exchange of
energy among the degrees of freedom of the particle, so that the particle escapes from the trap \cite{Chirikov1959} (Pedagogy: section 4.2 of \cite{Lichtenberg2013} and section 5.2.2 of \cite{Elskens2003}); in modern language, this corresponds to a large scale chaotic motion of the particle. This criterion is immediately applied successfully to the determination of the confinement threshold for experiments with plasmas of open mirror traps\footnote{References \cite{Chirikov1987a,Chirikov1987b,Chirikov1993} yield a broad account of particle confinement in magnetic traps.} \cite{Rodionov1959}. This simple to implement criterion becomes rapidly famous among physicists\footnote{An indication of the importance of this criterion is obtained when typing ``resonance overlap criterion" in Google Scholar: 209,000 references are obtained; \cite{Chirikov1979} is quoted 4,200 times according to the Web of Science.
Though not quoting Chirikov's seminal work, reference \cite{Rosenbluth1966} contributed to publicize the concept of resonance overlap too. In reality, the resonances supposed to overlap exist only in the limit where only one is present and the system is integrable. When the size of both trapping domains increases, a Poincar\'e map shows that chaos makes progressively the integrable separatrices fuzzy, which rules out any overlap. Furthermore, a perturbative calculation shows there is a mutual repulsion of the separatrices, which makes their overlap more difficult \cite{Escande1979}. Therefore, the resonance overlap criterion should be more adequately named a ``heteroclinic connection criterion" (see figure 1 of \cite{Elskens1993}).}, and especially after Chirikov's review paper\footnote{This paper brings also a wealth of information about Hamiltonian chaos which is very useful for physicists
of the eighties to get acquainted with chaos theory (``stochasticity theory" at that time).} \cite{Chirikov1979}. The resonant domain of a single wave and the overlap of the resonance domains of two waves can be observed in a traveling wave tube, a kind of noiseless beam-plasma system \cite{Doveil2005a}.  Rechester and Stix, when dealing with magnetic chaos due to weak asymmetry in a tokamak, use this criterion to compute the width of narrow chaotic (``stochastic") layers next to the separatrix of an integrable system when it is perturbed\footnote{Here again, Chirikov's authorship of the resonance overlap criterion is overlooked. First estimates of the width of chaotic layers are given in \cite{Zaslavsky1968,Zaslavsky1971} with the overlap criterion. Rechester and Stix' estimates are improved in \cite{Escande1982a}.} \cite{Rechester1976}.

This criterion is useful for systems with many degrees of freedom too. Indeed, it can be directly applied to determine the energy border for strong chaos in the Fermi-Pasta-Ulam system when only a few long wave modes are initially excited \cite{Chirikov1966,Chirikov1973}. In \cite{Escande1994} one computes the Gibbsian probability distribution of the overlap parameter $s$ corresponding to two nearby resonances of the Hamiltonian of a chain of rotators. Requiring the
support of this distribution to be above the threshold of large scale chaos, gives the right threshold in energy above which the Gibbsian estimate of the specific heat at constant volume agrees with the time average of the estimate given by the fluctuations of the kinetic energy: one has a self-consistent check of the validity of Gibbs calculus using the observable $s$!

\paragraph{Renormalization approach}
\label{RenAp}

At the end of the seventies, Doveil is working on ion acoustic waves in a multipole plasma device where these waves are dispersive. This leads him to study numerically the (chaotic) dynamics of ions when several waves are present\footnote{Computers at that time are slow enough for the successive points of the Poincar\'e map to come successively on the screen of a monitor. This makes obvious that chaotic motion is not really stochastic. In particular, when islands are present in the chaotic sea, one can see long phases where the
orbit looks as trapped in the corresponding resonances, a feature incompatible with a stochastic process. This is due to a self-similar structure producing strong correlations and incomplete chaos \cite{ZSUC:1993}.}. For simplicity he focusses on the two-wave case. The Poincar\'e map corresponding to the motion of one particle in two longitudinal (plasma) waves displays a sequence of resonance islands related to higher order nonlinear resonances which
 become explicit by canonical transformations\footnote{This set of islands has a signature, a ``devil's staircase", which is experimentally observable in a traveling wave tube \cite{Macor2005}. The cancellation of a set of islands can build a barrier to transport in the same device \cite{Chandre2005}.}. Following the philosophy of Chirikov's review paper \cite{Chirikov1979}, it is then tempting to apply the resonance overlap criterion to two neighboring such resonances. The criterion is easier to apply if the corresponding Hamiltonian is approximated by a simpler one... corresponding to the motion of one particle in two longitudinal waves. The passage from the initial two-wave Hamiltonian to the transformed one is similar to the transform of
 Kadanoff's block-spin renormalization group  \cite{Escande1981a,Escande1981b} (Pedagogy: section 4.5 of \cite{Lichtenberg2013} and section 5.4 of \cite{Elskens2003})... where Chirikov's criterion is absent! Its practical implementation for more general Hamiltonians is then described in \cite{Escande1984} and in sections 3.1 and 4.1 of the review paper \cite{Escande1985}. Appendix B of the latter reference shows how to derive a renormalization for any KAM torus trapped into a resonance island. A one parameter renormalization scheme is derived for ``stochastic layers" in \cite{Escande1982a}. All these schemes are approximate ones in a physicist sense: the approximations are not mathematically controlled.

Later on several mathematical works try and cope with this shortcoming. A way to make the 1981 renormalization scheme rigorous is indicated in \cite{MacKay1995}. The ideas proposed originally in \cite{Escande1981a,Escande1981b} lead to approximate renormalization transformations enabling a very precise determination of the threshold of break-up of invariant tori for Hamiltonian systems with two degrees of freedom \cite{Chandre2002}; these transformations are similar to the
transformation of Kadanoff's block-spin renormalization, in the sense that they combine a process of elimination and rescaling. In \cite{Chandre2002} the break-up of invariant tori proves to be a universal mechanism and the renormalization flow is precisely described. In 2004, Koch brings this type of approach to a complete rigorous proof \cite{Koch2004}: it is a computer assisted proof of the existence of a fixed point with non-trivial scaling for the break-up of golden mean KAM tori.

\paragraph{Other approaches}

For Hamiltonians with zero or one primary resonance, one cannot apply the resonance overlap criterion or the above renormalization approach. Reference \cite{Codaccioni1982} shows how to compute the threshold of large
scale chaos by using the blow-up of the width of chaotic layers\footnote{One of the considered cases is the polynomial H\'enon-Heiles Hamiltonian \cite{Henon1964}. However, the technique of \cite{Codaccioni1982} is unable to detect integrability. Indeed, it predicts a blow-up of the width of a chaotic layer also for the integrable Hamiltonian obtained from H\'enon-Heiles' one by changing a sign in its formula!} as computed in \cite{Rechester1976}.

In 2000, the study of the Poincar\'e map of magnetic field lines in a toroidal confinement configuration called ``reversed field pinch" germane to the tokamak is the occasion of a paradoxical discovery: chaos decreases when a magnetic
 perturbation increases, in contradiction with the prediction of the resonance overlap criterion! This phenomenon stems from a separatrix disappearance due to an inverse saddle-node bifurcation \cite{Escande2000}, a process likely to occur in many Hamiltonian systems\footnote{It occurred in the discharges at high current of the RFX-mod reversed field pinch, first when stimulated by a modulated edge toroidal field \cite{Lorenzini2008}, and then spontaneously \cite{Lorenzini2009}.}.

\subsubsection{Working with maps}
\label{Wwm}

In the seventies, Greene is interested in the nature of magnetic field lines in stellarators, and in their corresponding return area-preserving map. He naturally focusses on the simplest example of such maps, the standard map, which is very easy to iterate on computers of that time. At the end of the 70's, in the same way as it is natural to focus on higher order nonlinear resonances in a Hamiltonian description, it is natural to focus on periodic orbits with a long period in
area-preserving maps. These periodic orbits correspond to O-points and X-points of resonance islands. While studying the stable periodic orbits approximating a given KAM torus when truncating the continuous fraction expansion of its winding number at high order, Greene notes they become unstable when the KAM torus breaks up. This leads him to his famous ``residue criterion" which provides a method for calculating, to very high accuracy, the parameter value for the destruction of the last torus\footnote{Four theorems show that large parts of this criterion have a firm
foundation, but not all cases have been analyzed: for instance, can a non-smooth circle have residues going to infinity \cite{MacKay1992}? If so, then one cannot infer from residues going to infinity that there is not a circle.} \cite{Greene1979} (Pedagogy: section 3.2a of \cite{Lichtenberg2013}). Defining the threshold of large scale chaos in a given domain of phase space means finding the threshold of break-up of the most robust KAM torus. The continued
fraction expansion of their winding number is found numerically to have a special form exhibited in \cite{Greene1986}\footnote{Greene's criterion originates from stellarator studies. In 1984 it feedbacks on them by enabling the derivation of Hamiltonian systems that are for all practical purposes integrable, with the particular application to design stellarators with almost non-chaotic magnetic fields \cite{Hanson1984}.}.

Greene's work is placed in a renormalization group setting by MacKay, then his student \cite{MacKay1983}. This work is closely related to the approximate renormalization described above. This triggers a dialog between the latter
renormalization and the rigorous one under the auspices of Greene's criterion during almost a decade \cite{Schmidt1980,Doveil1981,Doveil1982,Schmidt1982,Escande1982c,Mehr1984,Mackay1988a}. The existence of a fixed point with a non-trivial scaling for MacKay's renormalization is finally proved in 2010 \cite{Arioli2010}.

Reference \cite{MacKay1989} derives a simple criterion for non-existence of invariant tori. When applied to the Hamiltonian describing the motion of a particle in the field of two waves of section \ref{RenAp}, it gives results in close agreement with those of Greene's residue method.

Till now we considered maps such that the period of the motion on a torus is a monotonic function of the action of the torus: they are twist maps. We now deal with non-twist systems where the period goes through an extremum on a given torus; in such systems, the overlap criterion fails and KAM theorem cannot be applied. They are introduced in 1984 by Howard, motivated by multifrequency electron-cyclotron-resonance heating in plasmas \cite{Howard1984}. He derives an accurate
analytic reconnection threshold of the approximate separatrices of the pairs of islands corresponding to actions symmetrical with respect to that of the extremum period. Motivated by the study of magnetic chaos in systems with reversed shear configurations, del-Castillo-Negrete and  Morrison propose a prototype map called the standard non-twist map, and present a detailed renormalization group study of the non-twist transition to chaos \cite{Non_twist_0,Non_twist_2,Non_twist_1} : there is a new universality class in this transition. This stimulates a series of rigorous mathematical results: in \cite{Delshams2000}, the proof of persistence of critical circles and a partial justification of Greene's criterion, as generalized in \cite{Non_twist_2}. In \cite{Gonzalez-Enriquez2014}, the bifurcations of KAM tori are studied by using the classification of critical points of a potential as provided by Singularity Theory. This approach is applicable to both the close-to-integrable case and the far from integrable case whenever a bifurcation of invariant tori has been detected numerically, but the system is not necessarily written as a perturbation of an integrable one.

\subsubsection{Finite time mappings for Hamiltonian flows}
\label{CanonTransMap}
In view of the many techniques which can be used for area preserving maps, it is interesting to construct a finite time mapping corresponding to a given Hamiltonian flow. In the nineties, motivated by various plasma physics issues, Abdullaev develops to this end his mathematically rigorous ``mapping method'' based on Hamilton-Jacobi theory and classical perturbation theory which works for
Hamiltonians that are the sum of an integrable part and of a small perturbation \cite{Abdullaev_99,Abdullaev_02} (See \cite{Abdullaev:2006,Abdullaev:2014} for a systematic description of the method). This method can be used to find infinitely many periodic orbits of a perturbed Hamiltonian system, and not only primary ones (see Sec.~7.2.2 in \cite{Abdullaev:2014}). It can also be used as a method of symplectic integration, with an accuracy controlled by the product of the perturbation parameter and of the mapping time step \cite{Abdullaev_02}. The method enables to derive
the canonical versions of mappings widely used in the theory of chaotic Hamiltonian systems and of their applications (see
\cite{Abdullaev_04a,Abdullaev:2006,Abdullaev_07} and references therein).

Using this method, the canonical separatrix mapping describing the dynamics near a separatrix is derived in
\cite{Abdullaev_04b,Abdullaev_05}, while keeping the canonical variables of the corresponding Hamiltonian, an improvement over Chirikov's separatrix mapping \cite{Chirikov1979}. The new mapping is consistent with the rescaling invariance described in section \ref{rescaling_invariance}, and enables the derivation of analytical formulas for the stable and unstable manifolds of the saddle point
\cite{Abdullaev_14,Abdullaev:2014}.

\subsection{Chaotic transport}
\label{ChTr}

In Hamiltonian systems with 1.5 or 2 degrees-of-freedom, when KAM tori break up in a given domain of phase space, chaotic transport sets in. In the sixties, the lack of mathematical results enabling the description of this transport and its quantification, leads plasma physicists to assume that chaotic motion is a stochastic one close to a Brownian motion and has a diffusive nature. However, as shown below, this is not the whole story!

\subsubsection{Quasilinear diffusion}
\label{QLDiff}

\paragraph{As an Ansatz}

In 1966, Rosenbluth, Sagdeev, and Taylor are interested in the destruction of magnetic field surfaces by magnetic field irregularities \cite{Rosenbluth1966}. They state that if there is resonance overlap, ``then a Brownian motion of flux lines and rapid destruction of surfaces results". They go to the action-angle variables of the unperturbed magnetic field, and write a Liouville equation for
field lines which is very similar to the Vlasov equation. Since the magnetic modes play a role for field lines analogous to that of Langmuir waves for electrons in a plasma, they use the quasilinear estimate of transport corresponding to the latter case. This estimate, introduced in 1961 \cite{RF}, had been made popular in 1962 by two papers on the bump-on-tail instability published in two successive issues of \textit{Nuclear Fusion} \cite{Vedenov1962,Drummond1962} (Pedagogy: section 5.4 of \cite{Lichtenberg2013} and section 6.8.1 of \cite{Elskens2003}). Then, quasilinear
estimates are made systematically for chaotic transport for almost three decades without questioning its validity, except for the standard map which exhibits a diffusive behavior with a diffusion constant oscillating as a function of the control parameter of the map about the quasilinear value\footnote{If the orbits inside accelerating islands are also taken into account, one finds a superdiffusive transport \cite{Benkadda1997}. Transport becomes diffusive by adding some noise to the mapping \cite{Karney1982}.} \cite{Chirikov1979,Rechester1980,Rechester1981,Meiss1983}. However, in 1998 the diffusion properties of the Chirikov-Taylor standard map are shown to be nonuniversal in the
framework of the wave-particle interaction, because this map corresponds to a spectrum of waves
whose initial phases are all correlated \cite{Benisti1998b}. For the sawtooth map, the dynamics is found diffusive for most of the integer values of the perturbation parameter \cite{Cary1981b}. This is done by calculating the characteristic functions, and the joint probabilities of the map \cite{Cary1981c}.

\paragraph{Quasilinear or not?}

For a chaotic motion, the perturbative approach used in the original derivation of the quasilinear equations cannot be justified. Therefore, the quasilinear description might not be correct to describe the saturation of the bump-on-tail instability; its inconsistency is shown in \cite{Laval1983a,Laval1983b}. In 1984,  Laval and  Pesme propose a new Ansatz to substitute the quasilinear one, and predict that the velocity diffusion coefficient should be renormalized by a factor 2.2
during the saturation of the instability \cite{Laval1984}. This motivates Tsunoda, Doveil, and Malmberg to perform an experiment with an electron beam in a traveling wave tube, in order to avoid the noise present in beam-plasma systems \cite{Tsunoda1987a,Tsunoda1987b,TSU}. This experiment comes with a surprising result: quasilinear predictions look right, while quasilinear assumptions are proved to be completely wrong. Indeed no renormalization is measured, but mode-mode coupling is not negligible at all. This sets the issue: is there a rigorous way to justify quasilinear estimates for chaotic dynamics?

This issue is first tackled by the author of this paper by considering the self-consistent motion of a finite number of waves and particles corresponding to the beam-plasma problem (see the first paragraph of section \ref{Scdy}).
However, the motion of a particle in a prescribed spectrum of waves is mysterious too and deserves a thorough study. For uncorrelated phases, it is
natural to expect the diffusion coefficient to converge to its quasilinear estimate from below when the resonance overlap of the waves increases. In 1990 numerical simulations of the motion of one particle in a spectrum of waves performed by Verga come with a surprising result: for intermediate values of the overlap, the diffusion coefficient exceeds its quasilinear value by a factor about 2.5 \cite{Cary1990}.

\paragraph{Diffusion or not?}

This result triggers a series of works aiming at understanding whether the diffusion picture makes sense and when the quasilinear estimate is correct. In 1997, B\'enisti shows numerically that the diffusion picture is right, provided
adequate averages are performed on the dynamics \cite{Benisti1997} (see also section 6.2 of \cite{Elskens2003}, and \cite{Escande2007,Escande2008}); however, this picture is wrong if one averages only over the initial positions of particles with the same initial velocity, in contrast with the 1966 Brownian Ansatz: chaotic does not mean Brownian! This motivates Elskens to look for mathematical proofs using probabilistic techniques, and leads him to two results: individual diffusion and particle decorrelation are proved for the
dynamics of a particle in a set of waves with the same wavenumber and integer frequencies if their electric field is gaussian \cite{Elskens2010}, or if their phases have enough randomness \cite{Elskens2012}. The intuitive reason for the validity of the diffusive picture is given in \cite{Benisti1997}: it is due to the locality in velocity of the wave-particle interaction, which makes the particle to be acted upon by a series of uncorrelated dynamics when experiencing large scale chaos. This
locality of the wave-particle interaction is rigorously proved by B\'enisti in \cite{Benisti1998a}. On taking into account that the effect of two phases on the dynamics is felt only after a long time when there is strong resonance overlap, it can be approximately proved that the diffusion coefficient is larger than quasilinear, but converges to this value when the resonance overlap goes to infinity \cite{Benisti1997,Escande2002b} (see also \cite{Escande2002a,Escande2003}, and Pedagogy: section 6.8.2 of \cite{Elskens2003}).

For the advection of particles in drift waves or in 2-dimensional turbulence, care must be exerted when trying to define a corresponding diffusive transport. Then one must define the Kubo number $K$ which is the ratio of the correlation time
of the stochastic potential as seen by the moving object to the (nonlinear) time where the dynamics is strongly perturbed by this potential (trapping time, chaos separation time, Lyapunov time, ...) \cite{ref6,ref12}. At a given time, the potential has troughs and peaks. If the potential is frozen, particles bounce in these troughs and peaks. When  $K \ll 1$, the particles typically run only along a small arc of the trapped orbits of the instantaneous potential during a correlation time (see figure 1a of \cite{Escande2007}). During the next correlation time they perform a
similar motion in a potential completely uncorrelated with the previous one. These uncorrelated random steps yield a 2 dimensional Brownian motion with a diffusion coefficient which can be computed with a quasilinear estimate.

\subsubsection{Diffusion with trajectory trapping}

If $K \gg 1$, a quasi-adiabatic picture works: the particles make a lot of bounces before the potential changes its topography (see figure 1b of \cite{Escande2007}). The change of topography forces particles to jump to a nearby trough or peak. The successive jumps produce a random walk whose order of magnitude of the corresponding diffusion coefficient can be easily computed \cite{ref6,ref12,Escande2007,Escande2008}. In a series of works, Vlad and coworkers clarify the issue of diffusion
with trajectory trapping. The just described simple picture is almost correct for a Gaussian spatial correlation function of the potential \cite{ref12}. More generally, the frozen potential displays as well \textquotedblleft{}roads\textquotedblright{} crossing the whole chaotic
domain. This enables long flights in the dynamics that bring some dependence upon $K$ in the estimate for the diffusion coefficient. The correct calculation of this coefficient is a much harder task. To this end, one may group together the trajectories with a high degree of similarity, and one starts the averaging
procedure over these groups. This yields the decorrelation trajectory method \cite{Vlad1998} and the nested subensemble approach \cite{ref12}. These techniques are extensively used for the study of the transport in magnetically confined plasmas (see \cite{Vlad2013} for a recent set of references), and for the study of astrophysical plasmas \cite{Vlad2014} and of fluids \cite{Vlad2015}. Reference \cite{Vlad1998} computes numerically the diffusion coefficient of particles in a spectum of waves scaling like $k^{-3}$, and for Kubo numbers up to $2 10^5$. For large $K$'s, due to trajectory trapping, the scaling of the diffusion coefficient with $K$ is less than $K^1$ corresponding to the Bohm scaling. For $1 < K < 10^4$, the results agree with the percolation scaling $K^{0.7}$ \cite{Isichenko}, but the scaling $K^{0.64}$ provided by the decorrelation trajectory method fits better the data in the whole domain $1 < K < 2 10^5$. A diffusion coefficient proportional to $K^{2/3}$, better than the percolation scaling, is obtained by a simple random-walk model using the concept of Hamiltonian pseudochaos, i.e. random non-chaotic dynamics with zero Lyapunov exponents \cite{Milovanov2009}.

Some turbulent plasmas may be modeled by integrable Hamiltonian systems subjected to non-smooth perturbations. Then, chaotic transport occurs at any small magnitude of perturbation. The profile of the diffusion coefficient in the unperturbed action is found to have a fractal--like structure with a reduced or vanishing value of the coefficient near low-order rational tori \cite{Abdullaev_11b}.

\subsubsection{Pinch velocity}

In reality, when the diffusive picture is correct, there is a pinch or dynamic friction part on top of the diffusive part, and the correct model is the Fokker-Planck equation \cite{Escande2007,Escande2008}. For the advection of particles in drift waves or in 2-dimensional turbulence, the direction of this pinch part depends on $K$ \cite{ref17}.

\subsubsection{Rescaling invariance of chaotic transport in chaotic layers}
\label{rescaling_invariance}

Consider a one-dimensional Hamiltonian which is the sum of an integrable part displaying a hyperbolic fixed point $X$ and of a time-periodic perturbation with amplitude $\epsilon$. Its phase-space near $X$ turns out to
be invariant with respect to a rescaling of the conjugate coordinates along the eigenvectors of $X$, of $\epsilon$, and of the phase of the perturbation. In the middle of the nineties, Abdullaev and Zaslavsky show it numerically \cite{AbdullaevZaslavsky_94,ZaslavskyAbdullaev_95}, and prove it rigorously \cite{AbdullaevZaslavsky_95b,Abdullaev_97} (see also
\cite{Abdullaev_00,Abdullaev:2006}). Since the motion slows down near a saddle point, a particle spends relatively large time intervals there.
Therefore, the transport of particles in a narrow stochastic layer about the separatrix related to $X$ is mainly determined by the structure of this layer near this point. It turns out that the statistical properties of chaotic transport are periodic (or
quasiperiodic) functions of $\log\epsilon$ \cite{Abdullaev_00} (see also
\cite{AbdullaevSpatschek_99,Abdullaev:2006}).

\subsubsection{Transport through cantori}

In the eighties it becomes clear among plasma physicists that chaotic transport is intrinsically more intricate than a diffusion, especially if one considers a single realization of the physical system of interest. In particular, it may be strongly inhomogeneous in phase space due to localized objects restricting it: the cantori described now.

When a KAM torus breaks up, it becomes a Cantor set called a cantorus \cite{Aubry1978,Percival1980} (Pedagogy: section IIB of \cite{Meiss2015}). In 1984, MacKay, Meiss, and Percival show that a cantorus is a leaky barrier for chaotic orbits, and that the flux through the cantorus between two successive iterates of the Poincar\'e map can be computed as the area of a turnstile built in a way similar to homoclinic lobes for X-points \cite{MacKay1984a} (Pedagogy: section IIA of \cite{Meiss2015}). The above-described renormalization theories for KAM tori provide a critical exponent for this area \cite{MacKay1984a}. The latter can be obtained
from the actions of homoclinic orbits \cite{MacKay1987}. A new description of transport in a chaotic domain can be obtained through Markov models combining the fluxes through the discrete set of the most important noble cantori \cite{MacKay1984b,Meiss1985,Meiss1986}. Such models also enable computing the power law temporal decay of correlations and lifetimes \cite{Hanson1985,Meiss1986} first noted in \cite{Chirikov1981b,Karney1983}. All these ideas turn out to
be very useful in the next decades, as explained in \cite{Meiss2015}. An approach to barriers in a chaotic domain motivated by plasma physics consists in defining approximately invariant circles\footnote{This approach leads in 2008 to the definition of temperature contours for chaotic magnetic fields \cite{Hudson2008}. This work quotes a series of studies by non plasma physicists which led to it, with the introduction of ghost surfaces in particular.} \cite{Dewar1992}.

The Lyapunov exponent measures the mean rate of divergence of nearby orbits inside a chaotic domain. It gives a rough estimate of the decay rate of the exponential part of the correlation functions, which is important in several plasma physics problems \cite{Grebogi1981}. It is generally computed numerically, but analytical estimates are available for some mappings (see \cite{Rechester1979}, and sections 5.2 and 6.3 of \cite{Chirikov1979} where it is called Krylov-Kolmogorov-Sinai entropy). It can be analytically computed for the motion of a particle in a broad spectrum of waves with a large amplitude (see section 6.8.2 of \cite{Elskens2003}).

\subsubsection{Symbolic dynamics for chaotic layers}
\label{Sdcl}

A striking regularity is present in the time series of the long chaotic orbits of the standard map that are in a stochastic layer surrounding a single island and bounded by two KAM tori: the radial coordinate of the moving point oscillates for a certain time in a region adjacent to an island chain, then jumps suddenly to another basin, where it remains for a random time, etc... This behavior can be
 adequately modelled by a Continuous Time Random Walk (CTRW) \cite{Balescu1997}. The issue of the diffusion of magnetic field lines in a tokamak leads to reconsider it in \cite{Misguich1998}. The associated time series can be described in terms of an algorithm based on a symbolic dynamics. A computer program enables a completely automatic measurement of the waiting times and of the transition probabilities of the CTRW, and therefore the analysis of arbitrary long time series.

\subsubsection{Transport in low shear or shearless systems}

As explained at the end of section \ref{Wwm}, the study of magnetic chaos in systems with reversed shear configurations motivated the introduction of the standard non-twist map. On varying the control parameter of this map above the break-up of the shearless curve, it is found that transport develops very slowly, because of structures with high stickiness giving rise to an effective barrier near the broken shearless curve \cite{Szezech2012}.

Internal transport barriers in toroidal pinches (tokamak and reversed field pinch) are favored by low or vanishing magnetic shear \cite{del-Castillo-Negrete1992}. This leads Firpo into the study of corresponding Hamiltonian models for the magnetic field lines, which brings conclusions with a general bearing. Indeed, low shear is shown to have a dual impact: away from resonances, it induces a drastic enhancement of the resilience to chaotic perturbations and decreases chaotic transport; close to low-order rationals, the opposite occurs \cite{Firpo2011}.

\section{Quasi order and order}
\label{Qord}
\subsection{Adiabatic theory}
\label{AdTh}

When dealing with configurations for the magnetic confinement of charged particles, one often finds that the motion of a particle in such a configuration has multiple scales. For instance, section \ref{ResOver} considered the case of magnetic mirror traps where a particle has a fast Larmor rotation and slow oscillations
along the lines of force. If the dynamics is not in a regime of large scale chaos, it is natural to take advantage of the time scale separation to describe the motion. This leads to tractable analytical calculations if the fast degree of freedom is nearly periodic compared to the slow one: one makes a (classical) adiabatic theory of the motion. In reality, the adiabatic ideas carry over to some non strictly adiabatic cases: this is neo-adiabatic theory. The applications of these ideas are now described.

\subsubsection{Classical adiabatic theory}
\label{NsAd}

Classical adiabatic theory is formalized in 1937 \cite{Krylov1936}. In the 1950s and early 1960s, Kruskal is working on asymptotics and on the preservation or destruction of magnetic flux surfaces. His unpublished work motivates Lenard and Gardner to develop a theory of adiabatic invariance to all orders \cite{Lenard1959,Gardner1959}. He then develops a Hamiltonian version of adiabatic theory \cite{Kruskal1962} where adiabatic invariants are related to proper action
variables. This technique is used to second order in \cite{Northrop1966,McNamara1967}. However its implementation is tedious, which prompts the use of other techniques: first the Poisson bracket technique with a methodological contribution from plasma physicists \cite{McNamara1967}, and then the powerful Lie transform method with two important methodological contributions from plasma physicists in 1976: one by Dewar \cite{Dewar1976}, and one by Dragt and Finn \cite{Dragt1976}\footnote{Cary's tutorial paper \cite{Cary1981a} provides a unifying view on Lie
 transform perturbation theory for Hamiltonian systems together with important applications in plasma physics.} (Pedagogy: section 2.3 of \cite{Lichtenberg2013}). Adiabatic motion in plasma physics is also a source of inspiration for pure mathematicians, as can be seen in \cite{Arnold1963b} which deals, in particular, with magnetic traps, and quotes Kruskal's work in his section devoted to adiabatic invariants.

\subsubsection{Neo-adiabatic theory}
\label{SiAd}

Several problems in plasma physics where there is a slow variation of the system of interest cannot be addressed by classical adiabatic theory. This is in particular the case when this slow variation induces a transition from trapped to passing orbits in magnetic configurations of magnetic fusion or of the magnetosphere. Then orbits cross a separatrix. Since the period of a motion diverges on a
separatrix, whatever slow be the evolution of the mechanical system, classical adiabatic theory breaks down to describe this crossing. However, it turns out that one can still take advantage of a separation of time scales for most crossing orbits: those which do not stick too long to the X-point. In 1986, four (groups of) authors come up with the calculation of the change of adiabatic invariant due to separatrix crossing: \cite{Neishtadt1986}, \cite{Hannay1986}, \cite{Vasilev1986},
 and a group of plasma physicists \cite{Tennyson1986,Cary1986}. The approaches are very similar (except for the third paper) and constitute what is now called neo-adiabatic theory\footnote{\cite{Bazzani2014} provides an extensive list of papers on neo-adiabatic theory. An early work already gave the solution in the case of a pendulum \cite{Timofeev1978}.}. The theory provides also explicit formulas for the trapping probabilities in a resonance region.

\subsubsection{Adiabatic description of Hamiltonian chaos}
\label{AddesHc}

Section \ref{QLDiff} considered the case of diffusive transport of a particle in strongly overlapping longitudinal waves. The diffusive picture was justified by the locality in velocity of the wave-particle interaction. This locality is quantified by a width in velocity which grows with the overlap parameter. Then, the diffusive picture is justified if this width is much smaller than the range of the phase velocities of the waves with strong resonance overlap. In the opposite case, the locality in velocity of the wave-particle interaction corresponds to a motion where
 the trapping time in the frozen potential of all waves is much smaller than the time scale of variation of this potential (this corresponds to the case of a large Kubo number introduced in section \ref{QLDiff}). At a given time, the frozen potential displays one or more separatrices which are pulsating with time. This issue is of interest to plasma physicists\footnote{In 1997, the understanding
  of adiabatic chaos leads to finding a way of mitigating its effects, such as in the work on omnigenous stellarators, viz. stellarators where all orbits are confined \cite{Cary1997}.}.

The simplest case corresponds to a single pulsating separatrix, as occurs for the dynamics of a nonlinear pendulum in a slowly modulated gravity field. Numerical simulations reveal that the domain swept by the slowly pulsating separatrix in the Poincar\'e map looks like a chaotic sea where no island is visible \cite{Menyuk1985,Elskens1993} (Pedagogy: section 1 of \cite{Elskens1993} and section 5.5.2 of \cite{Elskens2003}). As a result one might think the limit of infinite overlap to correspond to some ``pure" chaos. A fact pushing in this direction is a theorem stating that, in the domain swept by the slowly pulsating separatrix, the homoclinic tangle is
tight\footnote{When resonance overlap diminishes, at some moment the heteroclinic intersection between manifolds of the two resonances vanishes. This occurs at a threshold approximately given  by the resonance overlap criterion if the two resonances are not too different in size and wavelength \cite{Escande1981b,Escande1985}.} \cite{Elskens1991}. However another theorem tells the total area covered by small islands in the same domain generally decreases when the slowness of the system increases, but remains finite for symmetric frozen potentials \cite{Neishtadt1997}. This shows that, when taking at random initial conditions in the apparently smooth chaotic sea of the motion of a nonlinear pendulum in a slowly modulated gravity field, there is a finite probability to find a regular orbit: chaotic does not mean stochastic\footnote{Furthermore, the presence of the small islands induces the apparent intermittent trapping of chaotic orbits in a way analogous to what described in the first footnote of paragraph \ref{RenAp}.}! This also shows that chaos is not pure at all, and that the numerical simulation
of orbits may provide a misleading information\footnote{If the mathematical model is thought as the approximation of a true physical system, the dynamics of the latter undergoes actually perturbations like noise. These perturbations are likely to smear out the many minuscule islands of \cite{Neishtadt1997}. Then, the above numerical simulation gives the right physical picture. This sets the important issue of the structural stability of mathematical models when embedded into more realistic ones: numerical simulations might be more realistic than the mathematical model they approximate!}. In the adiabatic limit, successive
separatrix crossings are not independent, which significantly affects transport \cite{Bruhwiler1989,Cary1989}. However the separation of nearby orbits is intuitive, since two such orbits may be separated when coming close to the X-point, one staying untrapped and the other one becoming trapped. The transition from stochastic diffusion in a large set of waves to slow chaos associated to a pulsating separatrix can be detected experimentally in a traveling wave tube \cite{Doveil2011}.

Adiabatic invariants of slowly varying Hamiltonian systems occur not only for almost periodic orbits, but also for chaotic orbits that wander ergodically over the energy surface of the system.  This type of invariant is important in statistical mechanics of many-body systems, but is also invoked by Lovelace \cite{Lovelace1979} for single particle dynamics in the context of beam-plasma equilibrium and stability. This leads Ott to consider the general question of how well these approximate constants are preserved.  Using multiple time scale techniques and adopting ideas from quasilinear theory, he shows, among other results, that the error in these invariants is much larger than the one for almost periodic orbits \cite{Ott1979}.

\subsubsection{Separatrix crossing for non-slowly varying dynamics}

In 2015, B\'enisti shows that for the rapidly varying dynamics of a particle in a sinusoidal wave with a large exponential growth and a small initial amplitude, one can still describe separatrix crossing \cite{action}. This is done in two steps. First, a perturbative analysis in the wave amplitude provides the change in action up to the time when the action remains nearly stationary after trapping. Then, adiabatic theory is used to describe the subsequent evolution of the orbit:
the perturbative and adiabatic descriptions are matched. The method can be generalized to non-sinusoidal potentials and to waves that do not simply grow exponentially in time\footnote{This method can also derive the particles' distribution function. This allows to easily calculate the nonlinear response of a cold beam to an electrostatic wave \cite{action}.}. This technique works for particles with a high enough initial velocity. Lower velocities can be described by neo-adiabatic theory \cite{action}.

\subsection{Chimeras}
\label{Chim}

Fast ions in a fusion reactor can excite many types of waves, and in particular the energetic particle mode (EPM) which can produce avalanches of such ions. The coherent nonlinear behaviour of an EPM can be described by the complex Ginzburg-Landau equation (equation (2) of \cite{Zonca2006}), which describes a vast array of phenomena including nonlinear waves, second-order phase transitions, Rayleigh-B\'enard convection and superconductivity. In 2014, Sethia and Sen find out that this equation gives chimera states \cite{Sethia2014}. These states represent a spontaneous breakup of a population of identical oscillators that are identically coupled, into subpopulations displaying synchronized and desynchronized behavior. Till then, they had been found to exist in weakly coupled systems and with some form of nonlocal coupling between the oscillators. The new result shows that neither of these conditions is essential for their existence.

\subsection{Solitons and solitary waves}
\label{SoTh}

Year 1965 brings a landmark in soliton\footnote{There is a wealth of textbooks on solitons; the Wikipedia article may be a good pedagogical startpoint.} theory: the discovery by Zabusky and Kruskal \cite{Zabusky1965} that Korteweg-de Vries\footnote{The Korteweg-de Vries equation describes
long-wavelength water waves and ion-acoustic waves in plasmas.} (KdV) solitary waves survive collision. At that time computers enable to compute the collision of two solitary waves (their existence was proved in 1895). The outcome is surprising: after the collision, both solitary waves recover their initial shapes. Two years after, the solution is provided by Gardner, Greene, Kruskal, and Miura, with the inverse scattering transform\footnote{Pedagogy: the Wikipedia entry with this title.} \cite{Gardner1967,Gardner1974} which is a key technique of soliton theory with a lot of applications beyond plasma physics. During the period 1970 to 1982 many experiments on solitons in plasmas were performed (see \cite{Lonngren1983} for a review).

In 1970, Kadomtsev and Petviashvili \cite{Kadomtsev1970} introduce the first partial differential equation generating solitons in two space dimensions, as a model to study the evolution of long ion-acoustic waves of small amplitude propagating in plasmas under the effect of long transverse perturbations. This equation is a universal model (a kind of normal form) for nonlinear, weakly dispersive waves in an anisotropic medium \cite{Biondini2008}.

In 1971,  Rogister shows that the evolution of small-amplitude nonlinear Alfv\'en waves propagating quasiparallel with respect to the background magnetic field is governed by an equation called the derivative nonlinear Schr\"odinger equation \cite{Rogister1971}. In 1978, Kaup and Newell prove that this equation has solitonic solutions too \cite{Kaup1978}.

The effect of energy dissipation on solitary waves is important for plasma physics. This starts to be examined in 1969 by Ott and Sudan. Using a multiple time scale technique, they analyze the damping of solitary waves, first for the case of ion acoustic waves \cite{Ott1969}, and subsequently for very general types of damping mechanisms \cite{Ott1970}.

In 1983, Kaw, Sen and Valeo initiate an approach to the evolution of interacting waves in plasma, which consists in reducing the initial problem stated in terms of partial differential equations into a nonlinear Hamiltonian form with two degrees of freedom only \cite{Kaw1983}. This permits the identification and classification of a rich variety of nonlinear solutions, in particular solitary waves, but also periodic and stochastic solutions \cite{Kaw1983,Kaw1985,Kaw1992,Bisai1996}. Reference \cite{Kaw1983} elicits also interest in the nonlinear dynamics community because of the curious square root form of the nonlinear potential which seemed to make the system fully integrable. Eventually the system was shown to be non-integrable by an application of Ziglin's theorem \cite{Grammaticos1987}.

In 2012, Sen's research work on solitons in plasmas brings an interesting spin-off. A genetic programming search for equations sharing the KdV solitary wave solution uncovers a KdV-like advection-dispersion equation and an infinite dimensional family of equations with this property \cite{Sen2012}.

\section{High dimensional Hamiltonian dynamics}
\label{hdH}
\subsection{Self-consistent dynamics}
\label{Scdy}

\subsubsection{Finite dimension}

Self-consistent dynamics is at the heart of the complex behavior exhibited by plasmas. Indeed, the electric and magnetic fields determining the motion of charged particles are generated by the particles themselves. As a result, the particle dynamics in a plasma corresponds to a coupled Hamiltonian dynamical system with a very large number of degrees of freedom. This occurs in particular in the wave-particle interaction. References \cite{Escande1991,Antoni1998} and chapter 2 of \cite{Elskens2003} derive the Hamiltonian describing the self-consistent evolution of $N'$ tail particles and $M$ longitudinal waves, by starting from the $N$-body description of a one-dimensional plasma with a finite length (Pedagogy: section 2.1 of \cite{Elskens2003}). As explained in section \ref{QLDiff}, this derivation is motivated by a finite dimensional approach to the explanation of the surprising experimental result of \cite{Tsunoda1987a,Tsunoda1987b,TSU}.

The case with many waves displays a phenomenon known in fluid dynamics as ``depletion of nonlinearity": if the tail distribution function is a plateau in both velocity and space, which occurs at the saturation of the bump-on-tail instability introduced in section \ref{QLDiff}, the self-consistency is quenched, since the particles are not able to modify the wave amplitudes (see section 2.2 of \cite{Besse2011}). Therefore, their dynamics, even when strongly chaotic, is the 1.5
 degree-of-freedom one corresponding to the motion in a prescribed spectrum of waves. Depending on this spectrum, the diffusion coefficient may be quasilinear or not, as recalled in section \ref{QLDiff}. This contradicts previous works trying to prove the validity of quasilinear theory \cite{LD1,LD2,Escande2002b,Elskens2003} and the ``turbulent trapping'' Ansatz aiming at the contrary \cite{Laval1984} (see section 2.3 of \cite{Besse2011}). This brings a first element toward understanding the surprising experimental result recalled above.

The derivation for the single wave case ($M=1$) was first provided in two seminal papers \cite{Onishchenko1970,ONeil1971}. It was also provided in \cite{Tennyson1994,Mean_Field_Models_1,Crawford1999} by starting with kinetic descriptions. The single wave case is a paradigm for the interaction of particles with collective degrees of freedom: electrostatic instabilities \cite{Crawford1999}, interaction of vortices with finite-velocity flow in hydrodynamics \cite{Mean_Field_Models_2,Mean_Field_Models_3,Mean_Field_Models_4}. These references study the corresponding transition to chaos and formation of coherent
structures in phase space: order and chaos coexist, a typical feature of complex systems, already indicated by the existence of chimeras. The $M=1$ case also brings the possibility of analytical calculations which are impossible in the full $N$-body description of the plasma. Indeed, Gibbs statistical mechanics can be derived in this case, revealing a second order phase transition associated with the Landau damping regime: for nonequilibrium initial data with warm particles, above a critical initial wave intensity, thermodynamics predicts a finite wave amplitude in the limit $N \rightarrow \infty$; below it, the equilibrium amplitude vanishes \cite{Firpo2000}.

\subsubsection{Infinite dimension}

Due to its interaction with resonant particles, a longitudinal wave in a thermal plasma experiences a non-dissipative damping discovered by Lev Landau in 1946. This effect is of paramount importance in plasma physics. In a Vlasovian approach, it is understood as the consequence of a phase-mixing effect of a continuum of linear modes, called van Kampen modes\footnote{\cite{Morrison2000h} solves the dynamics of this continuum in the context of Hamiltonian systems theory, by a canonical transformation to action-angle
variables for this infinite degree-of-freedom system.}. However, one may wonder whether nonlinear effects do not destroy these linear modes and the corresponding phase mixing. Proving
the innocuity of nonlinear effects is the equivalent of deriving a KAM theorem for a continuous system (the Vlasov-Poisson one), a tour de force which partly earned Villani the 2010 Fields medal \cite{MV,Villani2014}. This is a major contribution of plasma physics to the nonlinear dynamics of continuous systems\footnote{The physical interpretation of Landau damping is subtle, but can be made completely intuitive by using the above finite dimensional self-consistent approach which shows that it is due to the synchronization of quasi-resonant particles with the wave \cite{Escande1996,Elskens2003};  the synchronization is confirmed experimentally \cite{Doveil2005b}. Within this approach, proving Landau damping in a nonlinear context requires the standard KAM theorem only.}.

When the electron distribution function is gradually changed, for instance by adding a bump in the tail of the distribution, an instability may appear in the Vlasov-Poisson system describing a spatially uniform plasma. For such a problem, center manifold theory cannot be used because of the existence of a continuum of modes. In order to cope with this problem, in 1989, Crawford and Hislop use the method of spectral deformation, a technique till then used in the theory of Schr\"odinger operators in quantum mechanics. They derive equations for the nonlinear
evolution of electrostatic waves by extending the method to the full nonlinear Vlasov equation, without making the standard assumptions of weak nonlinearity and separated time scales \cite{Crawford1989}. In 1994, Crawford overcomes the absence of a finite-dimensional center manifold in this problem, by restricting his analysis to initial conditions where only the unstable mode is initially excited, so that it is not one component of an arbitrary fluctuation. This way, he can treat the Vlasov equation perturbatively, and shows that for a plasma with a
neutralizing background of ions, the instability saturates at an amplitude scaling like the square of the growth rate, in heavy contrast with the traditional scaling like the square root. Furthermore, all orders contribute to the saturation, which implies that the equilibrium may be approached in a non monotonic way \cite{Crawford1994}.

As already indicated in section \ref{Chim}, the self-consistent interaction of fast ions with waves is an important topic for burning magnetic fusion plasmas. An analogy mapping the previous velocity distribution to the radial distribution of particles in a toroidal plasma provides a description of the corresponding limit case of a uniform plasma by the continuous limit of the above finite dimensional self-consistent model \cite{Zonca2015}. In 2015, Zonca and coworkers show that the more general description of the magnetic fusion case is provided by an analogue
of Dyson's equation in quantum field theory, describing particle transport due to emission and reabsorption of toroidal symmetry breaking perturbations, called ``phase space zonal structures" (PSZS), by analogy with the meso-scale configuration space patterns spontaneously generated by drift-wave turbulence. The relevant dynamics corresponds to the non-adiabatic (chaotic) case where the particle trapping time is not short with respect to the characteristic time for the nonlinear evolution of PSZS. For a non uniform plasma, there is a convective amplification of wave packets as avalanches leading to the secular transport of particles over large radial scales inside the toroidal plasma. This physics has analogies with the ``super-radiance" regime in free-electron lasers \cite{Zonca2015}.

Avalanches are also present in gyrokinetic numerical simulations of tokamak plasmas, which correspond to a reduced Vlasovian description of such plasmas taking advantage of the fact that the magnetic moment of particles is an adiabatic invariant. Avalanches correspond to a description of transport at strong variance with that of a chaotic transport due to turbulent waves. This description is germane with self-organized criticality, the kind of self-organization at work in sandpiles. This view progressively emerged in the last two decades \cite{Dendy1997,Dendy2007,Sanchez2015}.

\subsection{Noncanonical Hamiltonian theory}

The standard Hamiltonian description of physical systems uses canonical variables. This makes this description uneasy for systems that are written in terms of noncanonical variables like ideal fluid and systems with long range interactions when described by Vlasov equation. This motivates Greene and Morrison to introduce in 1980 noncanonical Poisson brackets for fluid systems \cite{Morrison1980}. Morrison
 brings this approach to full maturity in the following two decades (see \cite{Morrison1998} for a review). This approach leads in particular to the energy-Casimir criterion of nonlinear stability\footnote{This criterion was introduced first in \cite{Holm1985}, and in a series of papers where several plasma physicists are present (see note 42 of \cite{Morrison1998}).} (see section VI.B of \cite{Morrison1998}). Furthermore, linearly stable equilibria
 with negative energy modes are shown to be unstable when nonlinearity or dissipation is added (see section VI of \cite{Morrison1998} for a global discussion).

\section{Dissipative dynamics}
\label{disdyn}
While a wealth of plasma physics issues are naturally dealt with by a Hamiltonian approach, dissipative effects are important in plasmas too\footnote{See \cite{Greiner1993,Klinger1995,Mausbach1999} for a simple experimental proof of various phenomena like period doubling bifurcations and intermittency in low-pressure thermionic discharges. Pedagogy: chapter 7 of \cite{Lichtenberg2013}.}. This is the case for nonlinear wave coupling. In particular, for the dynamics of an unstable wave coupled nonlinearly to two lower frequency damped waves \cite{Vyshkind1976}. In 1980, Wersinger, Finn, and Ott demonstrate via the Poincar\'e section technique that this dynamics is well
described by a one-dimensional map with a quadratic maximum\footnote{This type of map exhibits period doubling cascades.}, which implies chaos with a strange attractor \cite{Wersinger1980}. This is the first explicit numerical demonstration of the applicability of such a map from physically motivated differential equations. As a sequel of this work, numerical techniques are used to compute for the first time the fractal dimension of strange attractors in several examples.
These dimensions are then compared with the predictions from the Kaplan-Yorke conjecture, which relates an attractor's fractal dimension to its Lyapunov exponents, thus providing important early confirmation for the conjecture \cite{Russell1980}.

In 1982, Bussac publishes an analytical method which accounts for the main features of the asymptotic solution of this coupled wave dynamics for the nonlinear decay of a coherent unstable wave into its subharmonic \cite{Bussac1982b}. This dynamics is shown to be close to a Hamiltonian one. The map providing the mismatch to the corresponding constant energy at each crossing of the Poincar\'e surface of section exhibits period doubling, and an explicit equation is obtained for the chaotic
attractor. The same type of nonlinear wave coupling can lead to another path to chaos called type I intermittency. When the control parameter is varied, the transition to chaos is abrupt, but the quantities which measure the amount of chaos smoothly vary with the control parameter, which implies a continuous character for such a transition \cite{Bussac1982c}. Reference \cite{Bussac1982a} yields a
complete panorama of the dynamics of the three-wave system involving again a one dimensional map which is analytically derivable from the original system of differential equations.

In 2003, Firpo shows that the description of the early nonlinear regime of the resistive $m=n=1$ mode of the tokamak can be done by assuming that the perturbation retains the form of the linearly unstable eigenmode, which leads to a Landau nonlinear stability equation \cite{Firpo2003,Firpo2004,Firpo2005}. The difficulty and novelty of this analysis comes from the existence of an inner critical
layer whose position is not fixed, in contrast with those occurring next to walls in fluid \cite{Stuart1958} and other plasma physics problems \cite{Dahlburg1998} whose solutions follow similar paths. In the much simpler problem of the nonlinear tearing mode where the position of the inner critical layer is fixed, Ottaviani and the author of the present paper show in 2004 that the perturbation can
be written as a linear unstable eigenmode plus a higher-order correction $\psi$ whose radial width scales like the linear magnetic island width. This yields a linear differential equation for $\psi$, which can be solved readily \cite{Escande2004}.

\section{Quantum chaos}
\label{Qcha}
In 1979, the issue of the chaotic motion of geometric optics rays in plasmas motivates McDonald and  Kaufman to study the quantum chaos provided by the two-dimensional Helmholtz equation with ``stadium'' boundary. In contrast to the clustering found for a separable equation, the eigenvalue separations have a Wigner distribution, characteristic of a random Hamiltonian \cite{McDonald1979}.

The same year, a work involving Chirikov considers the quantum kicked rotator, which is the quantized version of the standard map. In the regime of strong chaos, the rotator energy and the squared number of excited quantum levels appear to grow diffusively in time, as in the corresponding classical system. Only up to a break time though \cite{Casati1979},
after which the quantum energy excitation is suppressed while the classical one continues to diffuse. This time is proportional to the classical diffusion rate \cite{Chirikov1981a}.

However, even classically small noise
can induce quantum decoherence, thus eliminating the quantum saturation of diffusion, and restoring perpetual classical diffusion \cite{Ott1984}. This type of behavior has been experimentally investigated in atomic physics experiments \cite{Arndt1991}.

Multi-photon ionization of hydrogen can be treated by classical theory when the initial quantum number is large and the photon energy is small. Then ionization corresponds to the chaotic motion obtained for high photon intensities. However, the quantum ionization threshold may be higher than the threshold of classical chaos, because classical chaos is suppressed by quantum effects when the phase-space area escaping through \emph{classical cantori} each period of the electric field is smaller than Planck's constant \cite{Mackay1988b}.

\section{Natural extensions}
 \label{Pex}

There are natural extensions to this work. In particular, those due to intertwined contributions of plasma physicists with fluid dynamicists about vortex structures (see chapter 6 of \cite{Horton2002}). Also about the generalized Lagrangian mean, a formalism to unambiguously split a motion into a mean part and a nonlinear oscillatory part, where plasma physicists provided important contributions \cite{Frieman1960,Dewar1970}. There is also the use of non-neutral plasmas for the experimental study of vortex dynamics in two-dimensional hydrodynamics \cite{Malmberg1992,Durkin2000}.

\section{Conclusion}

The main part of this story is borne out of two constraints: the need for a theoretical description of the magnetic confinement of thermonuclear plasmas and the insufficient corresponding mathematical knowledge. However its birth would not have been possible without the development of computer calculations.

Plasma physics is often considered as motivated by its applications. This is wrong, since plasmas belong in complex systems which are of interest on their own, and are a central topic of contemporary physics. This is right, provided one counts applications like its just recalled numerous contributions to chaos and nonlinear dynamics!

A histogram of the number of published papers quoted in this topical review as a function of time shows that there is a peak of the number of published papers per year in the 1980-1984  period, then a decrease to 60\% of this peak in the 1985 to 2004 period, and to 40\% in the 2005-2014 period, with no decrease when going from 2005-2009 to 2010-2014. Therefore, there still is a steady flow of contributions from plasma physics to chaos and nonlinear dynamics.

However, there is less excitement about these new results than in the 1980-1984  period where plasma physicists started to understand chaos. The 1980-1984 results recalled in this topical review bring a reassuring background to the present investigations, but most are not much used explicitly. Indeed, as to fusion theory, the development of massive computer simulations progressively revealed that the complexity of fusion plasmas displays a mixture of order and chaos, for instance in avalanches, that is not suggestive of an elementary quantitative description
(so far). Furthermore, numerical calculations are so handy, that it is more reliable to find out a threshold of bifurcation numerically, for instance a threshold of chaos, than to compute it by applying a low dimensional technique after painstaking uncontrolled approximations. The practical success of the simple-minded quasilinear approximation was another incentive to give up more sophisticated descriptions of chaotic transport. Works about nonlinear or chaotic plasma physics have been taking over: a whole series of books would be necessary to summarize their results! However a breakthrough in the fundamental analysis of complex plasma dynamics cannot be excluded...

I thank P. Diamond, U. Frisch, and Y. Pomeau for inciting me to study the topic of this topical review. I thank S. Abdullaev, D. B\'enisti, J. Cary, D. del-Castillo-Negrete, R. Dewar, F. Doveil, Y. Elskens, M.-C.~Firpo, U. Frisch, J. Krommes, R. MacKay, J. Meiss, J.T.~Mendon\c{c}a, J. Misguich,  P.  Morrison,  E. Ott, A. Sen, D. Shepelyansky, J. B. Taylor, M. Vlad, and F. Zonca for their initial inputs to this work and for successive suggestions. Also, Y. Elskens, R. MacKay, and P. Morrison who provided me extensive advice on my manuscript. I thank R. Dendy for suggestions and also for questions, which fed the concluding remarks, and L.~Cou\"edel for his help in managing the references.
\bibliographystyle{dcu}
\bibliography{bibliodominique7}

\end{document}